\documentclass[prb,aps,twocolumn,showpacs,groupedaddress,amsmath,amssymb]{revtex4-1}

\usepackage{graphicx}
\usepackage{dcolumn}
\usepackage{bm}
\usepackage{subfigure}
\usepackage{color}
\usepackage{verbatim}

\begin{document}

\title
{ 
Fractionalization, entanglement, and separation: \\
understanding the collective excitations in a spin-orbital chain
}

\author{Cheng-Chien Chen$^{1}$, Michel van Veenendaal$^{1, 2}$, Thomas P. Devereaux$^{3}$, Krzysztof Wohlfeld$^{3,4}$
}
\affiliation{$^{1}$Advanced Photon Source, Argonne National Laboratory, Argonne, Illinois 60439, USA}
\affiliation{$^{2}$Department of Physics, Northern Illinois University, De Kalb, Illinois 60115, USA}
\affiliation{$^{3}$Stanford Institute for Materials and Energy Sciences, SLAC National Laboratory and Stanford University, Menlo Park, California 94025, USA}
\affiliation{$^{4}$Institute of Theoretical Physics, Faculty of Physics, University of Warsaw, Pasteura 5, PL-02093 Warsaw, Poland}
\date{\today}

\begin{abstract}
Using a combined analytical and numerical approach, we study the collective spin and orbital excitations in a spin-orbital chain under a crystal field.
Irrespective of the crystal field strength, these excitations can be universally described by fractionalized fermions.
The fractionalization phenomenon persists and contrasts strikingly with the case of a spin chain, where fractionalized spinons cannot be individually observed but confined to form magnons in a strong magnetic field.
In the spin-orbital chain, each of the fractional quasiparticles carries both spin and orbital quantum numbers, and the two variables are always entangled in the collective excitations.
Our result further shows that the recently reported separation phenomenon occurs when crystal fields fully polarize the orbital degrees of freedom.
In this case, however, the spinon and orbiton dynamics are decoupled solely because of a redefinition of the spin and orbital quantum numbers.
\end{abstract}

\pacs{75.25.Dk, 71.10.Fd, 75.10.Pq, 78.70.-g}

\maketitle

\section{Introduction}
Strong correlation effects can lead to intriguing emergent phenomena, such as the creation of quasiparticles like phonons and magnons.
Being long-lived objects with a well-defined energy-momentum dispersion,
these ``new particles'' exist as eigenstates of the low-energy effective Hamiltonian.
Their statistics and quantum numbers, however, can be exotic and different from those of the constituent particles.
One well-known example is the spin chain discussed below.

\subsection{A reference system: spin chain}
{\it Fractional excitation.}
In a spin $S=1/2$ chain, when the spins show ferromagnetic (FM) alignments,
the elementary excitations are $S=1$ spin-flip (magnon) excitations,
and the corresponding spectrum exhibits a sharp, single-magnon mode~\cite{Mueller1981, Brenig2009, Mourigal2013}. 
Naively, one might expect a similar scenario for the $S=1/2$ Heisenberg chain with nearest-neighbor (NN) antiferromagnetic (AF) interactions.
In that case, the ground state is ``almost" ordered with a slowly decreasing power-law AF correlation,
in which spins tend to form local SU(2) singlets with their neighbors.
However, instead of a magnon excitation, a spin flip creates two elementary excitations -- called spinons -- related to the formation of magnetic domain walls [Fig. 1(a)].
Each spinon carries half of the spin quantum number of a magnon~\cite{Faddeev1981} and no charge quantum number.
The phenomenon of carrying only a fraction of the quantum numbers from the underlying constituents is referred to as {\it fractionalization}~\cite{Heeger1988, Laughlin1999, Kivelson2002}.

{\it Spinon confinement in strong magnetic field.}
Spinons in an AF background are deconfined, as they can move away from each other spatially without costing extra energy.
The spectrum of a spin-flip excitation (creating two spinons) thereby develops an energy continuum~\cite{Mueller1981, Dender1997, Brenig2009, Mourigal2013}.
This is quite different when spin degeneracy is lifted by a magnetic field $H_z$.
For an $H_z$ exceeding the critical strength $H^{cr}_z$ that sustains a FM ground state,
spin excitation is {\it no longer fractional};
spinons cannot be individually observed but confined as magnons [Fig. 1(b)].
In this case, inelastic neutron or x-ray scattering experiments, which probe single spin-flip excitation, 
would measure only a sharp, single-magnon mode~\cite{Mourigal2013}.

{\it Spin-charge separation.}
Another exotic property of a spin chain is the potential separation of quantum numbers.
Upon doping a hole, another fractional elementary excitation called a holon appears.
Unlike spinons which carry spin 1/2 but no charge, holons carry spin 0 and charge $e$.
The spinon and holon are decoupled and propagate at different velocities, showing the separation of spin and charge quantum numbers carried respectively by two different fractionalized quasiparticles~\cite{Lieb1968, Luther1974, Kim1996, Kim2006, Moreno2013}.

\begin{figure}[t!]
\includegraphics[width=\columnwidth]{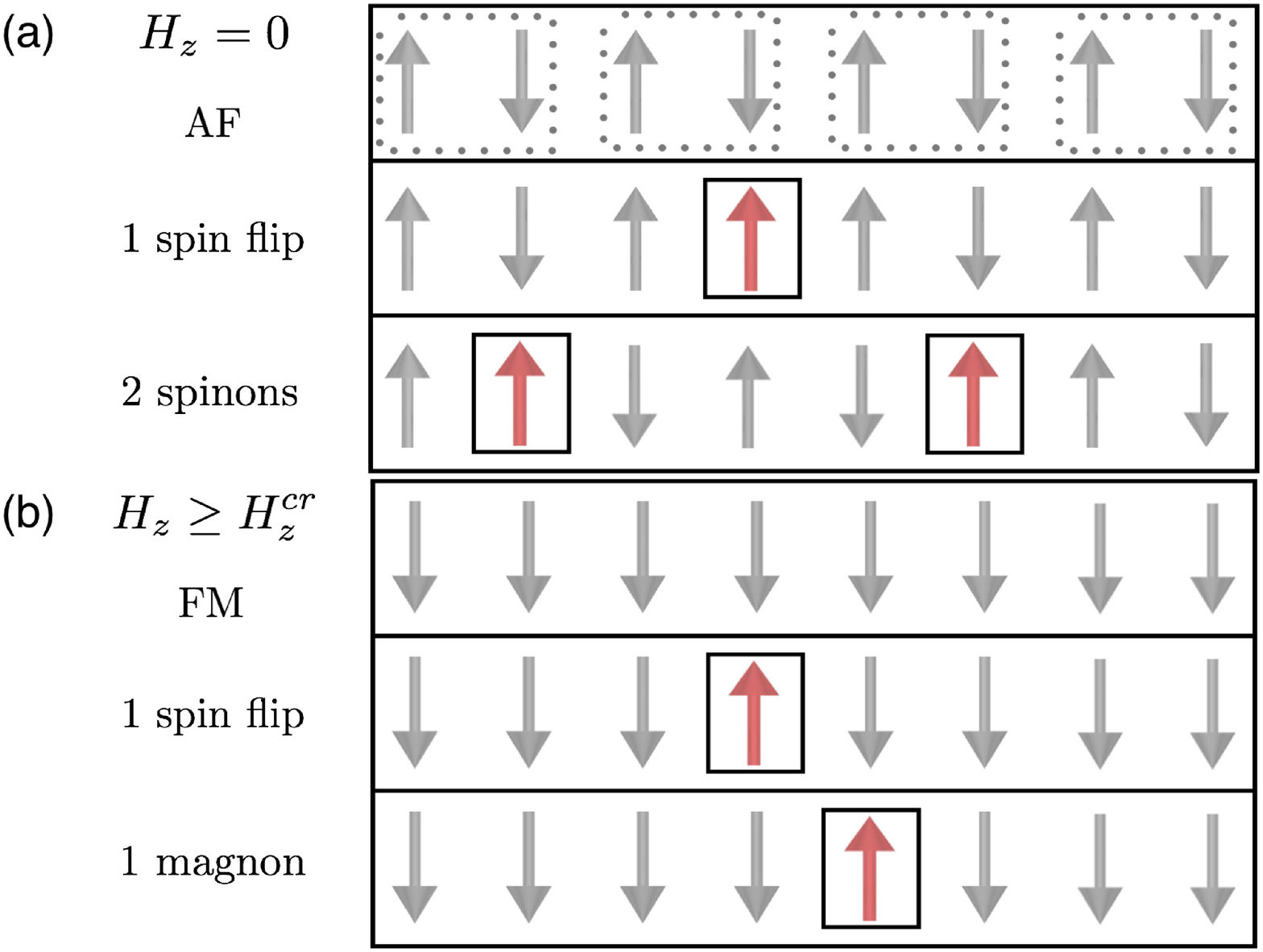}
\caption{
Illustrations showing collective excitations in a spin chain~\cite{Mueller1981, Brenig2009, Mourigal2013}:
(a) Without magnetic field ($H_z=0$), the ground state exhibits AF correlations (top row, showing only one fluctuating SU(2)-singlet configuration denoted by the dotted rectangle).
A spin flip (middle row) creates two {\it fractionalized} spinons (magnetic domain walls between parallel spins).
The spinons are deconfined and can propagate through two lattice sites via one exchange interaction (bottom row).
(b) For magnetic fields larger than the critical strength ($H_z \geq H_z^{cr}$), the ground state is ferromagnetic (top row).
A spin-flip excitation (middle row) {\it does not fractionalize}.
Instead, the domain walls are confined as a single magnon excitation when moving through the chain (bottom row).
}
\label{fig:scartoon}
\end{figure}

\subsection{A related system: spin-orbital chain}
In addition to spin, the orbital degrees of freedom play an important role in the low-energy physics of various correlated transition-metal compounds~\cite{Kugel1982,Tokura2000}.
Recent advance in resonant inelastic x-ray scattering (RIXS) now allows this technique to directly probe orbital excitations over almost the entire Brillouin zone~\cite{Ulrich2009, Ament2011, Schlappa2012, Benckiser2013}.
This has revived the studies of various model Hamiltonians with competing or cooperating  spin-orbital interactions~\cite{Wohlfeld2011, WenLong2012, Kumar2013, Wohlfeld2013, Brzezicki2013}.
One of the simplest models is the one-dimensional (1D) Kugel-Khomskii Hamiltonian~\cite{Kugel1982}:
\begin{equation}\label{eq:h}
\mathcal{H} = 4J \sum_{\langle i j \rangle} \Big( {\bf S}_i \cdot {\bf S}_j + \frac{1}{4} \Big) \Big( {\bf T}_i \cdot {\bf T}_j + \frac{1}{4} \Big) + E_z  \sum_i T^z_i.
\end{equation}
Here ${\bf S}_i$ (or ${\bf T}_i$) is an SU(2)-invariant spin (or pseudospin) 1/2 operator at site $i$, $\langle ij\rangle$ represents an NN pair, $J$ is the superexchange energy, and $E_z$ is the crystal field strength.
Such a model emerges in the strong coupling limit for a chain consisting of two orbitals per site expressed in terms of the pseudospin operator.
It also describes spin ladders with four-spin interactions~\cite{Momoi2003, Gritsev2004, Lecheminant2005}.

{\it SU(4) limit: fractionalization and entanglement.}
Without the crystal field term ($E_z=0$), Eq. (\ref{eq:h}) has an enlarged SU(4) symmetry~\cite{Zhang1998, Zhang1999, Frischmuth1999}.
The SU(4) spin-orbital model can be regarded as a generalization of the SU(2) spin chain to higher symmetry representation,
relevant to cold-atom measurements of SU($N$) antiferromagnets in optical lattices~\cite{Gorshkov2010, Bonnes2012, Messio2012}.
The model also serves as a good starting point to describe the spin and orbital properties of real condensed matter systems~\cite{Kugel2014}.

\begin{figure}[t!]
\includegraphics[width=\columnwidth]{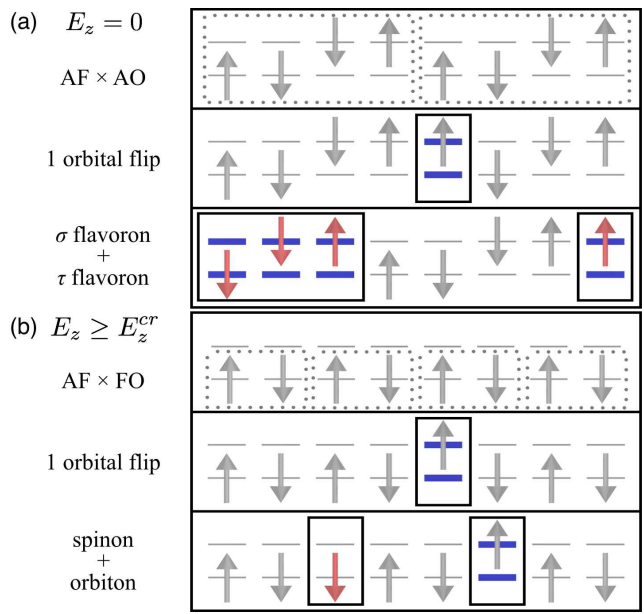}
\caption{
Illustrations showing collective excitations in the spin-orbital chain in two different limits described in the literature~\cite{Zhang1999, Wohlfeld2011}:
(a) Without crystal field ($E_z=0$),
the ground state exhibits AF$\times$AO correlations described by SU(4) singlets (top row, showing only one fluctuating configuration denoted by the dotted rectangle).
A spin (or orbital) flip predominantly {\it fractionalizes} into the $\sigma$ and $\tau$ flavorons~\cite{Zhang1999};
they are ``free" quasiparticles, carrying {\it entangled} spin and orbital quantum numbers (bottom row).
(b) In a strong crystal field ($E_z \geq E_z^{cr}$),
electrons occupy 
only the the lower-lying orbitals and show AF correlations (AF$\times$FO).
An orbital flip (middle row) was suggested to fractionalize into 
{\it separate} spinon and orbiton, which respectively carry only the spin and orbital quantum number (bottom row)~\cite{Wohlfeld2011}.
}
\label{fig:socartoon}
\end{figure}

Due to its high symmetry, the 1D SU(4) spin-orbital model is exactly solvable by a numerical Bethe Ansatz~\cite{Sutherland1975, Zhang1999}.
The ground state was found to show AF and alternating orbital (AO) correlations (AF$\times$AO) [Fig.~\ref{fig:socartoon}(a)],
which can be described as a superposition of SU(4) singlets~\cite{Zhang1998, Zhang1999, Frischmuth1999}.
In this case, a single spin or orbital flip {\it fractionalizes} into different ``flavorons" -- collective excitations carrying entangled fractional spin and orbital quantum numbers [Fig.~\ref{fig:socartoon}(a)]~\cite{Chen2007,WenLong2012, Lundgren2012}.

{\it Strong crystal field: fractionalization and separation.}
Although the SU(4) spin-orbital model has been applied previously to real systems~\cite{Arovas1995, Zhang1998, Pati1998},
its condensed matter realization is still limited.
Orbitals (unlike spins) can assume different shapes in real space, and thereby realistic effective orbital interactions usually have lower symmetries.
However, when $E_z \ge E^{cr}_z$ (where $E^{cr}_z$ is the critical field for inducing a ferro-orbital (FO) ground state),
Eq. (\ref{eq:h}) has been shown to capture successfully the spin-orbital physics in a number of quasi-1D cuprate materials~\cite{Wohlfeld2011, Wohlfeld2013}.
This success mainly relies on the relatively small energy scales of lower-symmetry interactions (originating from the Hund's coupling and orbital-dependent hoppings),
which do not qualitatively change the underlying physics in the limit of large $E_z$~\cite{Wohlfeld2011, Wohlfeld2013}.

When $E_z \ge E^{cr}_z$, the ground state is described by SU(2) spin singlets with AF correlations between electrons occuping only the lower orbitals (AF$\times$FO) [Fig.~\ref{fig:socartoon}(b)].
An orbital-flip excitation from such a state was suggested theoretically to {\it fractionalize} into a spinon and an orbiton~\cite{Wohlfeld2011, Wohlfeld2013}, which respectively carry only spin and orbital quantum numbers [Fig.~\ref{fig:socartoon}(b)].
This {\it spin-orbital separation} has been reported by recent RIXS measurements on Sr$_2$CuO$_3$~\cite{Schlappa2012}, CaCu$_2$O$_3$~\cite{Bisogni2015}, and SrCuO$_2$~\cite{privcomm_thorsten_schmitt},
and has lead to further investigation of similar phenomena in other spin-orbital models~\cite{Kumar2013, Brzezicki2013}.

\subsection{Aim and structure of the paper}
In this paper, we establish {\it a unified description} of collective excitations in the spin-orbital chain at the isotropic SU(4)-symmetric point {\it and} the anisotropic limit of large crystal field.
Although the excitations are fractionalized in both cases, their natures seem quite distinct:
The spin and orbital are entangled on one hand, while they are reported to be separate on the other.
Bridging the theoretical gap between the two limits is crucial to a comprehensive understanding of the fundamental physics revealed by spectroscopic measurements on spin-orbital systems.
Moreover, it is important to understand the persistent fractionalization phenomenon despite the presence of a strong crystal field. 
The latter contrasts strikingly with the case of a spin chain in a strong magnetic field, where fractionalized spinons cannot be individually observed but confined as magnons of integral quantum numbers.

To answer these questions, we thoroughly investigate the spin-orbital model in Eq. (\ref{eq:h}) for $E_z$ ranging from 0 to $E^{cr}_z$.
By a combined numerical and analytical study, we unambiguously show that collective excitations of the spin-orbital chain are always fractional,
carrying entangled spin and orbital quantum numbers. 
We also show that the recently reported separation phenomenon occurs when crystal fields fully polarize the orbital degrees of freedom.
In this case, however, the spinon and orbiton dynamics are decoupled solely because of a redefinition of the spin and orbital quantum numbers.
Our numerically unbiased, highly quantitative calculations using cluster perturbation theory with exact diagonalization (CPT+ED) provide new excitation spectra at intermediate crystal fields.
In addition, our analytical large-$N$ mean-field approach is applied for the first time to the spin-orbital model,
providing an intuitive physical understanding of the collective excitations in terms of particle-hole excitations in non-interacting fermionic bands.
{\it This approach achieves a unified description of the spin-orbital model for all values of $E_z$}.
 
The rest of the paper is organized as follows:
Section II presents the numerical CPT+ED spectra of the spin and orbital dynamical structure factors.
Section III introduces the large-$N$ mean-field theory of constrained fermions and the resulting compact supports of spin and orbital excitations.
Section IV focuses on an effective $t$-$J$ model description (which has previously lead to the suggestion of spin-orbital separation) and discusses its limitation.
Section V summarizes our findings with additional concluding remarks.
The Appendix provides further details of the CPT+ED calculations and discusses the mapping of Eq. (\ref{eq:h}) onto an effective $t$-$J$ model.

\section{Numerical results}
\label{sec:numerics}

We begin to study Eq. (\ref{eq:h}) by computing the transverse spin and orbital dynamical structure factors: 
\begin{align} \label{eq:spindsf}
S(q, \omega)=\frac{1}{\pi} \lim_{\eta \rightarrow 0} 
\Im \langle \psi | S^x_{q} 
\frac{1}{\omega + E_{\psi}  - \mathcal{{H}} -
i \eta } S^x_q | \psi \rangle,
\end{align}
\begin{align} \label{eq:orbitaldsf}
O(q, \omega)=\frac{1}{\pi} \lim_{\eta \rightarrow 0} 
\Im \langle \psi | T^x_{q} 
\frac{1}{\omega + E_{\psi}  - \mathcal{{H}} -
i \eta } T^x_q | \psi \rangle.
\end{align}
Here $| \psi \rangle$ is the ground state of $\mathcal{{H}}$ with energy $E_{\psi}$,
$S^x_q \equiv \sum_j e^{iqj} S^x_j /\sqrt{L}$ is the Fourier transform of the local spin operator (the same applies to $T^x_q$), and $L$ is the number of lattice sites.
The dynamical structure factor is related to the Fourier transform of a real-space correlation function and provides the energy-momentum dispersion relation of elementary spin or orbital excitation.

We first employ the numerical CPT+ED technique~\cite{Senechal2000, Ovchinnikov2010, Senechal2012,Maska_PRB_1998} to compute $S(q, \omega)$ and $O(q, \omega)$.
CPT is a quantum cluster approach~\cite{Maier2005} complementing finite-size ED simulations.
It can be regarded as an efficient interpolation scheme to obtain spectra with continuous momentum transfers.
The method reproduces several known exact results for the spin chain and the spin-orbital model at a quantitative level (see Appendix A).
We note that the main spectral features discussed below also can be identified by ED,
and our conclusion of a fractional nature in the spin-orbital chain does not depend on the CPT implementation.
However, the additional fine spectral details provided by CPT+ED greatly facilitate the comparison of our numerical and analytical results.
Further CPT+ED calculations are detailed in Appendix A.

Figure 3 displays the CPT+ED spectra at different $E_z$ for spin (left panels) and orbital (right panels) dynamical structure factors.
When $E_z=0$ (top panels), the ground state shows AF$\times$AO correlations described by SU(4) singlets without any orbital 
polarization ($T^z_{tot} \equiv\sum_i T^z_i /L=0$).
The spin and orbital spectra are identical, with gapless excitations at $q=0$, $\pi/2$, and $\pi$.
For $E_z\neq 0$, the spectra can exhibit incommensurate soft modes.
When half of the orbitals are polarized ($T^z_{tot}=1/4$, middle panels),
the zero-enery spin excitations shift away from $q=\pi/2$;
the orbital excitations remain gapless at $q=\pi/2$ but gapped at $q=0$, $\pi$ (see Sec. III for an intuitive understanding).
When $E_z=E_z^{cr}$, the orbitals are fully polarized ($T^z_{tot}=1/2$, bottom panels),
where electrons reside only in the lower-lying orbitals and show AF correlations.
The spin spectrum consists of the one-spinon branch and two-spinon continuum as those in a spin chain;
the orbital spectrum is identical to the hole-addition spectrum in a $t$-$J$ model (see Sec. IV)~\cite{Brunner2000, Wohlfeld2011}.

The above results agree with Bethe-ansatz and density matrix renormalization group (DMRG) calculations~\cite{Yamashita2000,Yu2000},
showing incommensurate soft modes under external fields and broad energy continua implying fractional elementary excitations.
However, it is difficult to obtain the spectral weight information with Bethe-ansatz solutions;
it is also challenging to converge the DMRG results in longer chains or higher energies due to the enlarged spin-orbital basis.
Besides $S(q,\omega)$ and $O(q,\omega)$, we further compute the simultaneous spin-orbital flip spectra $OS(q,\omega)$ [obtained by replacing $S^x_q$ with $S^x_q T^x_q$ in Eq. (\ref{eq:spindsf})].
For all values of $E_z$, we find that $OS(q,\omega)$ follow exactly the dispersion of $O(q,\omega)$.
This holds only because the exchange interaction in Eq. (\ref{eq:h}) retains an SU(4) symmetry.
A strong Hund's coupling $J_H$, for example, can lower the symmetry to SU(2)$\times$U(1),
and $OS(q,\omega)$ no longer tracks $O(q,\omega)$.
This feature could benchmark the role of $J_H$ in materials such as V$_2$O$_3$~\cite{Castellani1998}.

\begin{figure}[t!]
\includegraphics[width=\columnwidth]{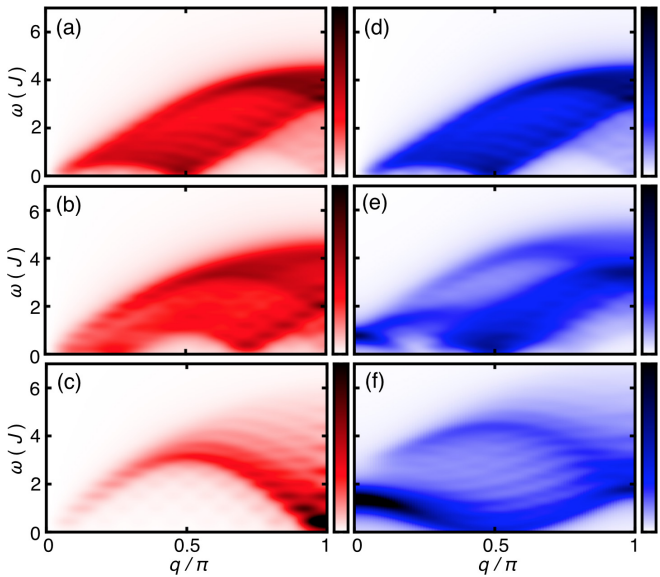}
\caption{
Dynamical structure factors for spin [(a)-(c)] and orbital [(d)-(f)] computed by CPT+ED at various $E_z$.
Top panels: $E_z=0$ with no orbital polarization;
Middle panels: $E_z\sim0.6E^{cr}_z$ with half polarized orbitals;
Bottom panels: $E_z=E^{cr}_z$ with fully polarized orbitals.
The false color white represents zero intensity, and black represents the maximal intensity (0.4 for (c) and 0.2 for the others).
The $L=16$ ED calculations are broadened by a 0.25$J$ Lorentzian.
The ripple structure resulting from CPT interpolation smooths and the overall spectral shape converges quickly with increasing $L$.
}
\label{fig:3}
\end{figure}

\section{Mean-field large-$N$ theory of constrained fermions}
\label{sec:analytics}

We next develop an analytical formalism to understand our numerical spectra.
We note that a direct mean-field decoupling of the spin and orbital variables in Eq. (\ref{eq:h}) fails to describe both its static and dynamic properties~\cite{Zhang1998, Zhang1999, Frischmuth1999, Wohlfeld2011}.
The two degrees of freedom show strong quantum entanglement and fluctuation~\cite{Chen2007,WenLong2012, Lundgren2012},
foreseeing the failure of a simple linear spin- or orbital-wave approximation~\cite{Wohlfeld2011}.
Here we use a different type of mean-field theory that was first developed for SU($N=2$) antiferromagnets~\cite{Baskaran1987}, and later generalized to large $N$~\cite{Affleck1988}.
This method concerns a {\it fermionic} representation of the exchange interaction, followed by a mean-field decoupling in terms of local valence bond singlets that preserve the SU($N$) symmetry of the problem~\cite{Arovas1988, Auerbach1994}.
As shown below, such approach captures the main features of the spin-orbital model even for $E_z\neq 0$.

\begin{figure}[t!]
\includegraphics[width=\columnwidth]{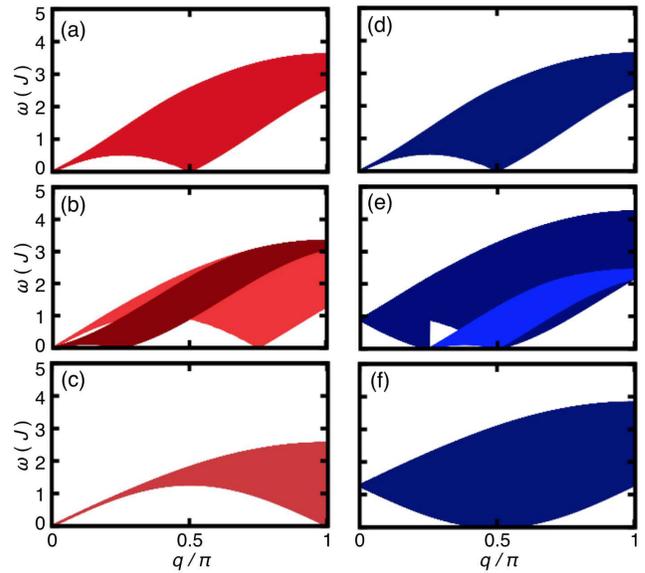}
\caption{
{\it Compact support} for spin [(a)-(c)] and orbital [(d)-(f)] spectra obtained by mean-field large-$N$ theory of constrained fermions at various $E_z$.
Top panels: $E_z=0$ with no orbital polarization; Middle panels:  $E_z = 2 \sqrt{2} J / \pi$  ( $\sim 0.7 E^{cr}_z$ in the mean-field picture) with 
half polarized orbitals; Bottom panels: $E_z=E^{cr}_z$ with fully polarized orbitals.
In (b), the darker (lighter) branch refers to spin-flip excitations of electrons in the upper (lower) orbitals.
In (e), the darker (lighter) branch refers to orbital-flip excitations caused by pseudospin raising (lowering) operators.
}
\label{fig:mfspectra}
\end{figure}

\begin{figure*}[t!]
\begin{minipage}{\textwidth}
\includegraphics[width=\textwidth]{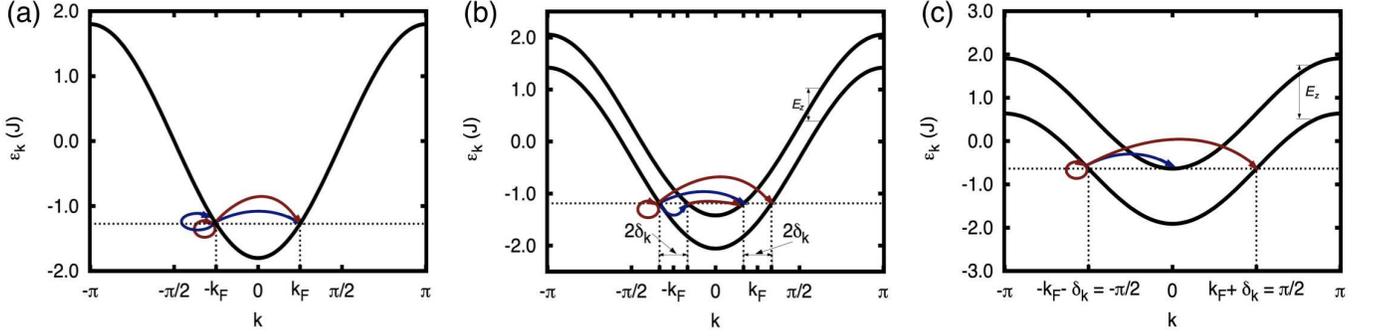}
\end{minipage}
\caption{
Evolution of the mean-field fermionic bands as a function of the crystal field at (a) $E_z=0$, (b) $E_z= E_z^{cr} / 2$, and (c) $E_z = E_z^{cr}$.
The collective spin and orbital excitations in the mean-field picture correspond to ``particle-hole" excitations of the constrained fermions across the Fermi surface (denoted by the dotted horizontal lines).
The energies of the $a$-orbital and $b$-orbital fermionic bands are separated by $E_z$,
and the allowed ``particle-hole" excitations change with the crystal field accordingly.
The thick arrows point to the allowed zero-energy spin (red) and orbital (blue) excitations.
}
\label{fig:mfbands}
\end{figure*}

We begin by noticing that {\it no charge fluctuation} is implicitly assumed in the spin-orbital model:
The system has exactly one particle per lattice site (quarter-filled), where double, triple and quadruple occupancies are prohibited.
We next express the spin and pseudospin operators of Eq. (\ref{eq:h}) in terms of the constrained fermion $c_{i \sigma}$ and Schwinger boson $p_{i \alpha}$ operators:
\begin{eqnarray}
S_i^+ = c_{i \uparrow}^\dagger c_{i \downarrow} \quad S_i^-= c_{i \downarrow}^\dagger c_{i \uparrow}, \\
T_i^+ = p_{i a}^\dagger p_{i b} \quad T_i^-= p_{i b}^\dagger p_{i a},
\end{eqnarray}
where $ \uparrow / \downarrow$ denotes one of the two spins $\sigma$,
and $a / b$ denotes one of the two orbitals $\alpha$.
[Note that one could also represent the spin (pseudospin) using the Schwinger boson (constrained
fermion) representation -- the choice is arbitrary].
These constrained fermions and Schwinger bosons fulfill the spin and pseudospin commutation relations provided that $ \sum_{ \sigma} c_{i \sigma}^\dagger c_{i \sigma}= 1$ and $\sum_{ \alpha} p^\dagger_{i \alpha} p_{i \alpha}= 1$.
We further define the constrained fermion carrying both spin {\it and} orbital quantum numbers:
\begin{equation}
{f}^\dagger_{i  \alpha \sigma} = c_{i \sigma}^\dagger p^\dagger_{i \alpha},
\end{equation}
with the constraint $\sum_{ \alpha \sigma} {f}^\dagger_{i, \alpha \sigma} {f}_{i, \alpha \sigma} = 1$ that follows from the constraints on the $c_{i\sigma}$ and ${p_{i\alpha}}$ occupation numbers.
By applying the above transformations, we finally arrive at the Hamiltonian of constrained fermions for the spin-orbital chain:
 \begin{align}\label{eq:hfermion}
\mathcal{H} = & -  J \sum_{\langle i j \rangle, \alpha \sigma, \alpha' \sigma'} 
\big( f^\dag_{i \alpha \sigma}  f_{j \alpha \sigma} + h.c. \big) 
\big( f^\dag_{j \alpha' \sigma'}  f_{i \alpha' \sigma'} + h.c. \big) \nonumber \\
&+ \frac12 E_z  \sum_{ i \sigma}  (f^\dag_{i a \sigma}  f_{i a \sigma}  - f^\dag_{i b \sigma}  f_{i b \sigma} ). 
 \end{align}

Here we note that both constrained fermion operators $c_{i\sigma}$ (due to the constraint of one fermion per site) and $f_{i \alpha \sigma}$ (due to its definition and also the constraint of one fermion per site) fulfill nonfermionic anticommutation rules, 
which is similar to the case of projected fermions in the $t$-$J$ model~\cite{Spalek2007}. 
However, these ``special" anticommutation relations will not be enforced in the following mean-field treatment, and $f_{i \alpha \sigma}$  will be referred to simply as fermionic operator.

At this stage, we perform a mean-field decoupling of the above four-fermion interactions in terms of local valence bond singlets,
$\chi_{i j} \equiv \sum_{\alpha, \sigma}( f^\dag_{i \alpha \sigma}  f_{j \alpha \sigma} + h.c. )$,
which preserve the SU(4) symmetry of the problem~\cite{Baskaran1987, Arovas1988}:
$\chi_{i j} \chi_{ji} \rightarrow ( \chi_{i j} \langle \chi_{j i}  \rangle  + \chi_{j i} \langle \chi_{i j}  \rangle) /2$.
The resulting mean-field Hamiltonian $\mathcal{H}_{\rm MF}$ reads
\begin{equation}\label{eq:hfermion}
\mathcal{H}_{\rm MF} = \sum_{k , \sigma} \big(  \varepsilon_{k a } f^\dag_{k a \sigma}  f_{k a \sigma} +  \varepsilon_{k b} f^\dag_{k b \sigma}  f_{k b \sigma} \big),
\end{equation}  
where $ \varepsilon_{k a  / b} = -  4 \sqrt{2} J \cos (\delta_k)  \cos (k) / \pi \mp E_z /2$,
with $ \delta_k = \arcsin [ E_z \pi /  (4 J)] / 2$ when $E_z <  4 J / \pi$, and $\delta_k = \pi / 4$ when $E_z \ge 4 J / \pi$~\cite{ft3,ft4,ft5}.
Determined by self-consistent mean-field equations,
$ \varepsilon_{k a  / b}$ represent the energies of two doubly degenerate fermionic bands separated by $E_z$ (see Fig. \ref{fig:mfbands}).
With the constraint of one fermion per site (fulfilled only on average at the mean-field level),
the bands are filled up to the respective Fermi momenta:
$\pm k_F \mp \delta_k $ and $ \pm k_F \pm \delta_k$,
where $k_F =\pi/4$ is the Fermi momentum at $E_z=0$, and $\delta_k $ (defined above) is an additional shift with nonzero $E_z$.

In this mean-field picture, collective spin and orbital excitations become ``particle-hole" excitations of the noninteracting constrained fermions across the Fermi level.
In particular, the spin spectra are related to excitations within the degenerate bands, while the orbital spectra are related to excitations between the nondegenerate bands.
The compact support (region where a function is nonzero) for the spin and orbital excitations can be computed respectively by 
 $\bar{S}(q, \omega) =  \sum_{k \in FS,  q+k \notin FS, \alpha,\sigma} \delta (\omega - \varepsilon_{q+k, \alpha \sigma } + \varepsilon_{k \alpha \bar{\sigma} })$,
where $\bar \sigma \equiv -\sigma$,
and $\bar{O}(q, \omega) =  \sum_{k \in FS,  q+k \notin FS, \sigma} \delta (\omega - \varepsilon_{q+k, a \sigma } + \varepsilon_{k b \sigma}) + \sum_{k \in FS,  q+k \notin FS} \delta (\omega - \varepsilon_{q+k, b \sigma} + \varepsilon_{k a \sigma})$.
The evolution of compact supports as a function of $E_z$ is shown in Fig.~\ref{fig:mfspectra}.

As seen in Figs.~3 and~4, the mean-field results agree well with the numerical simulations, revealing {\it inter alia} the shift in momentum of the zero-energy modes with $E_z$.
The origins of the zero-energy modes [{\it e.g.} at $q =0, \pi/2$ when $E_z=0$, and at $q= \pi/2$ (for orbital) or at $q=0, \pi$ (for spin) when $E_z = E_z^{cr}$] can be understood in terms of the allowed momenta for zero-energy particle-hole excitations between the occupied and unoccupied fermionic bands (see Fig. \ref{fig:mfbands}).
While the mean-field approach cannot account for the spectral intensity~\cite{Raczkowski2013},
it reproduces essentially the compact supports and the overall bandwidths for the spin and orbital dynamical structure factors.
Only the low-intensity branch of the numerical spectra are missing;
it originates from the $\omega$ flavoron~\cite{Zhang1999} and would involve four constrained fermions, which cannot be captured in the mean-field theory.

We last note that the above mean-field theory works better for the spin-orbital model with SU(4) exchange interaction than for the SU(2) spin chain.
More precisely, when $E_z=0$, the bandwidth of the spin excitation is $ \sqrt{2} \pi J$ in CPT+ED,
and $8 \sqrt{2} J / \pi$ in mean field (a factor of 1.23 difference).
When $E_z = E^{cr}_z$, the spin excitation bandwidth is $2 \pi J $ in CPT+ED, and $ 8 J  / \pi $ in mean field (a factor of 2.47 difference).
In the orbital spectra, however, the quantitative differences are smaller.
The extrapolated numerical critical field is $E^{cr}_z\sim1.38J$, only 1.08 times larger than the mean-field value $E_z^{cr} = 4 J / \pi$.
These results agree with Ref.~\onlinecite{Arovas1988}, showing that the mean-field approximation for SU($N$) antiferromagnets gradually improves as $N$ becomes larger~\cite{Arovas1988, Auerbach1994}.
In fact, the mean-field theory becomes exact for SU($N$) models when $N \rightarrow \infty$.
We also note that a {\it bosonic} theory usually works better for systems with long-range order, such as a fully spin-polarized chain or a square-lattice Heisenberg antiferromagnet~\cite{Auerbach1998}.
It thus cannot be applied to our case of a spin-orbital chain under an external crystal field, where only the orbital variables are polarized but no true spin long-range order develops.

\section{$t$-$J$ model description and its limitation}
\label{sec:tJ}

In this section, we discuss the use of an effective $t$-$J$ model to describe the spin-orbital chain, which has previously lead to the suggestion of spin-orbital separation~\cite{Wohlfeld2011, Wohlfeld2013, Schlappa2012, Bisogni2015}.
In particular, we show that the $t$-$J$ model description is valid only when $E_z \ge  E_z^{cr}$.
 
According to Appendix B, we can rewrite the spin-orbital model Eq. (\ref{eq:h}) as a bosonic $t$-$J$ model~\cite{Boninsegni2001}:
\begin{align} \label{eq:bosonictJ_v2}
\mathcal{H}^{\rm t-J} &=  J \sum_{\langle i, j \rangle, \sigma}  \big( b^\dagger_{i \sigma} b_{j \sigma} + h.c. \big) 
%+2J \sum_{\langle i, j \rangle} \Big( {\bf S}_{i b} {\bf S}_{ j b} + \frac{1}{4} \Big) \nonumber \\
+2J \sum_{\langle i, j \rangle} \Big( {\bf S}_{i} \cdot {\bf S}_{j} + \frac{1}{4} \Big) \nonumber \\
&+  E_z \sum_{i}  n_{b i},
\end{align}
where $b_{i\sigma}$ is a hard-core {\it boson} operator subject to the constraint $\sum_\sigma b^\dagger_{i\sigma} b_{i\sigma} \leq 1$.
In this case, electrons in the upper $a$ orbitals can be seen as holes in the spin background formed by electrons in the lower $b$ orbitals.

\begin{figure}[t!]
\includegraphics[width=\columnwidth]{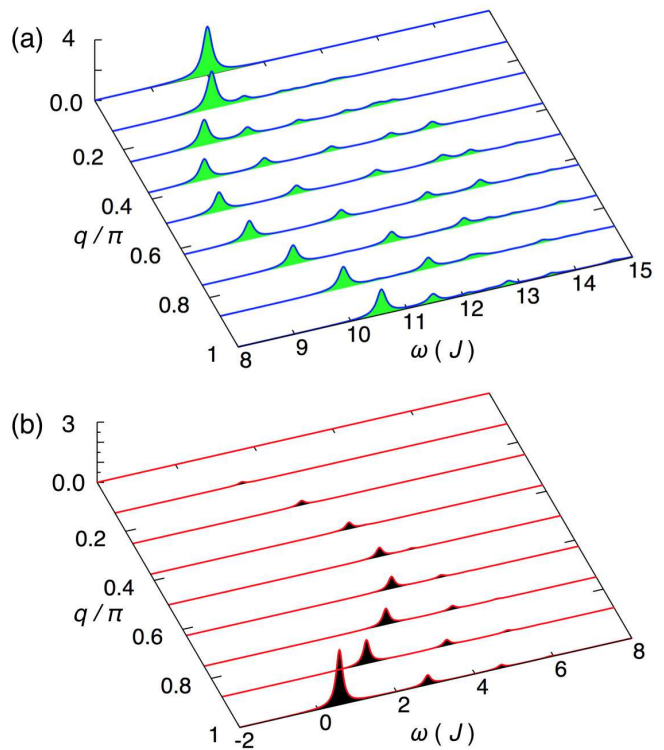} 
\caption{
Collective excitations computed by $L=16$ exact diagonalization at $E_z=10J (> E^{cr}_z)$~\cite{Wohlfeld2011},
where the spin-orbital model is ferro-orbitally ordered and mapped onto a half-filled $t-J$ model.
In (a), the orbital spectrum $O^{-+}(q,\omega)$ of the spin-orbital model (blue line) is the same as the hole spectral function $A(q,\omega)$ in the $t-J$ model (green area).
Note that $O^{+-}(q,\omega)=0$, since all lower orbital are occupied in the studied limit.
In (b), the spin spectrum in the spin-orbital model (red line) is identical to the spin spectrum in the $t-J$ model (black area). 
}
\label{fig:ChenbosonictJ}
\end{figure}

After the mapping, the orbital spectrum $O(q, \omega)=\frac{1}{4}[ O^{+-}(q,\omega)+O^{-+}(q,\omega)]$ for the spin-orbital model is equivalent to the hole spectral function $A(q,\omega)\equiv \sum_\sigma A_\sigma(q,\omega)$ in a half-filled $t$-$J$ model (with a factor of 2 difference in the spectral weight).
Here
\begin{eqnarray}
A_\sigma (q, \omega)&=&\frac{1}{\pi} \lim_{\eta \rightarrow 0} 
\Im \langle \psi | b^\dag_{q \sigma} 
\frac{1}{\omega + E_{\psi}  - \mathcal{H}^{\rm t-J}  -
i \eta } b_{q \sigma} | \psi \rangle,
\nonumber\\
O^{\mu\nu}(q, \omega)&=&\frac{1}{\pi} \lim_{\eta \rightarrow 0} 
\Im \langle \psi | T^\mu_{q} 
\frac{1}{\omega + E_{\psi}  - \mathcal{{H}} -
i \eta } T^\nu_q | \psi \rangle.
\nonumber\\
\end{eqnarray}
Since all the lower-lying orbitals are occupied when $E_z  \ge E_z^{cr}$, $O^{+-}(q,\omega)=0$ and $O^{-+}(q,\omega)=A(q,\omega)$.
Moreover, the spin spectra ${S}(q, \omega)$ for both models are identical.

To verify the above mapping, we compute the spectral functions for the two models using ED.
As shown in Fig. \ref{fig:ChenbosonictJ}, the results are in perfect agreements.
This demonstrates (for the first time) that the mapping to the effective bosonic $t$-$J$ model is completely correct when $E_z \geq E_z^{cr}$.
We also note that the physics of a bosonic $t$-$J$ model with $t>0$ is the same as that of a fermionic 
$t$-$J$ model with $t<0$. The latter follows from a Jordan-Wigner transformation of the spin-orbital Hamiltonian~\cite{Wohlfeld2011},
which is allowed when $E_z \geq E_z^{cr}$.

%which is allowed when $E_z \geq E_z^{cr}$. Thus, the above results also implicitly demonstrate the correctness of the mapping 
%from Ref. ~\onlinecite{Wohlfeld2011}. We also note that in this paper we decided to perform the mapping to the bosonic $t$-$J$ model, instead of using the mapping
%into the fermionic $t$-$J$ model from Ref.~\onlinecite{Wohlfeld2011}, since the earlier can more explicitly show why the mapping onto any effective $t$-$J$ model 
%fails when  $E_z < E_z^{cr}$, cf. Appendix B.

\begin{figure}[t!]
\includegraphics[width=\columnwidth]{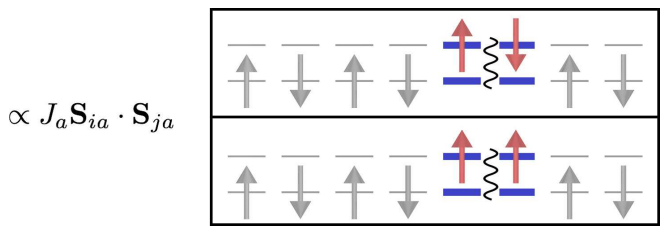}
\caption{
Illustrations showing the repulsive (attractive) interaction between two parallel (antiparallel) spins in the upper $a$ orbitals.
This interaction ($\propto J_a \mathbf{S}_{ia} \cdot \mathbf{S}_{ja}$) is neglected in the $t-J$ model approach, explaining why the mapping onto the $t-J$ model does not work for $E_z < E_z^{cr}$; see text for details.}
\label{fig:tJcartoon}
\end{figure}

When $E_z < E_z^{cr}$, however, more than one electron occupy the upper orbitals,
and the spin interaction between them cannot be neglected (see Fig.~ \ref{fig:tJcartoon} and Appendix B).
In this case, the spin-orbital model {\it cannot} be mapped to the bosonic $t-J$ model, and their spectra are distinct.

\section{Discussion and Conclusion}
\label{sec:discussion} 
%We first discuss the implications of our results.
The agreement between our numerical method and the analytical large-$N$ approach justifies the fermionic mean-field picture~\cite{Liuemail}.
(We note that in Ref.~\onlinecite{Kim1997}, a simple mean-field approach was also employed to justify spin-charge separation in 1D cuprates.)
The method immediately explains why $OS(q,\omega)$ follows the dispersion of $O(q,\omega)$:
a joint spin-orbital flip is described as particle-hole excitations between two bands with opposite spin and orbital quantum numbers. 
This produces the same ``topology" as that of pure orbital excitations, since the bands with different spin but the same orbital quantum numbers are degenerate (see Fig. 5).
This also enables us to arrive at the following {\it conclusions}:

{\it (i) Excitations are always fractional.}
We note that the mean-field constrained fermions are noninteracting ``good" quasiparticles,
and a single spin flip ($\Delta S=1$ excitation) can be understood as creating two independent fractionalized fermions (a particle and a hole) with quantum numbers $(S^z = 1/2, T^z = \alpha)$ and $(S^z=-1/2, T^z=\alpha)$, respectively.
Similarly, an orbital-flip excitation fractionalizes into the $(S^z = \sigma, T^z=1/2)$ and $(S^z =\sigma, T^z=-1/2)$ fermions.
Independent of $E_z$, fractionalization thereby always exists.
This contrasts strikingly with the case of a pure spin chain, where the elementary excitations are $S=1$ magnons when $H_z \geq H^{cr}_z$~\cite{Mueller1981, Brenig2009, Mourigal2013}.
In the spin-orbital model, $E_z$ acts only on the orbital variables and does not quench the quantum spin dynamics, allowing peculiar fractionalization even under large crystal fields.

{\it (ii) Spin and orbital are always entangled.}
According to the SU($N$) mean-field approach, irrespective of the values of $E_z$, the spin and orbital excitations can be described by fractionalized fermions (momentum eigenstates carrying both spin and orbital quantum numbers), and thereby the two degrees of freedom are always entangled.
This spin-orbital entanglement~\cite{Chen2007,WenLong2012, Lundgren2012, Oles2012}, 
however, has to be reconciled with the suggested spin-orbital separation for $E_z \ge E_z^{cr}$ in Refs.~\onlinecite{Wohlfeld2011, Wohlfeld2013, Schlappa2012, Bisogni2015}.
There, electrons in the lower orbitals {\it {are implicitly assumed}} to carry only spin but no orbital quantum number, whereas the single electron in the upper orbital carries only the orbital but no spin quantum number.
In this case, the spin-orbital model (having 4 degrees of freedom) can be mapped onto an effective $t$-$J$ model (having 3 degrees of freedom), as the spin degree of freedom of the electron in the upper orbital does not interact with other spins, leading to an emergent separation.
However, such separation can be regarded as a decoupling of the spinon and orbiton dynamics only because of a redefinition of the spin and orbital quantum numbers,
which is allowed only for describing a single orbital- or spin-flip excitation from the FO ground state.

In summary, using an unbiased, highly quantitative numerical technique and the large-$N$ mean-field theory of constrained fermions, 
we have formulated for the first time a unified framework to describe a spin-orbital chain in various regimes ranging from the isotropic SU(4)-symmetric point to the anisotropic limit of large crystal field.
The description based on SU($N$) mean-field theory provides an intuitive picture to understand the spin and orbital spectra in terms of particle-hole excitations between 
effective noninteracting fermionic bands. This study connects the novel physics predicted and observed in quasi-1D copper oxides~\cite{Wohlfeld2011, Wohlfeld2013, Schlappa2012, Bisogni2015} to the physics described by numerically exact Bethe-Ansatz solutions~\cite{Sutherland1975, Zhang1999,Yamashita2000},
and possibly to observables in optical lattice measurements with ultracold atoms~\cite{Gorshkov2010, Bonnes2012, Messio2012}.
Extensions of this study to lower symmetries and higher dimensions are interesting areas for future work.

%%[THIS PROBABLY IS NOT THE BEST TEXT, PLEASE THINK WHETHER WE CAN HAVE STH ELSE HERE;
%PERHAPS WE NEED SOMETHING ABOUT THE EXPERIMENT?; OR NOTHING?]
%These results also suggest that it will be interesting to investigate in the future the nature of spin and orbital excitations
%in the spin-orbital models in two or higher dimensions or in 1D with negative exchange constants. In these cases,
%there is no fractionalization but peculiar spin-orbital bound states showing spin-orbital entanglement 
%might appear in the absence of crystal field \cite{Brink1998, Shen2002, WenLong2012}. Therefore, in view of the presented results, 
%it would be interesting to verify whether such states would also not be affected when the, ubiquitous in the crystals, 
%external crystal field is taken into account. 

\section*{ACKNOWLEDGMENTS}
The authors acknowledge discussion with  Bruce Normand, Zheng-Xin Liu, Joseph Maciejko, Andrzej Ole\'s, Rajiv Singh, Tsez\'ar Seman, and Hong-Hao Tu.
C.C.C. is supported by the Aneesur Rahman Postdoctoral Fellowship at Argonne National Laboratory (ANL), operated by the U.S. Department of Energy (DOE) Contract No. DE-AC02-06CH11357.
M.v.V is supported by the DOE Office of Basic Energy Sciences (BES) Award No. DE-FG02-03ER46097 and the NIU Institute for Nanoscience, Engineering and Technology.
K. W. and T. P. D. acknowledge support from the DOE-BES Division of Materials Sciences and Engineering (DMSE) under Contract No. DE-AC02-76SF00515 (Stanford/SIMES). 
K. W. acknowledges support from the Polish National Science Center under Project No. 2012/04/A/ST3/00331.
K. W. is also grateful for support from the ANL X-ray Science Division Visitor Program.
The collaboration was supported by the  DOE-BES-DMSE Computational Materials Science Network program under Contract No. DE-FG02-08ER46540.
This work utilized computational resources at NERSC, supported by the U.S. DOE Contract No. DE-AC02-05CH11231.

\section*{APPENDIX A: CLUSTER PERTURBATION THEORY WITH EXACT DIAGONALIZATION}

Here we describe the numerical approach employed in this study, which involves an interpolation by cluster perturbation theory of the exact diagonalization spectra (CPT+ED)~\cite{Senechal2000, Ovchinnikov2010, Senechal2012}.
CPT is a quantum cluster approach~\cite{Maier2005} which can provide both dynamical and temporal information for quantum lattice models in the thermodynamic limit, thereby complementing finite-size ED simulations.
It also can be viewed as a simple and efficient method for obtaining spectra of continuous momentum transfer.
By benchmarking this method against exact Bethe-Ansatz solutions, we show that several known exact results for the spin chain and the spin-orbital model can be reproduced by CPT+ED at a quantitative level.

The CPT algorithm proceeds by (i) dividing the model lattice Hamiltonian into multiple identical, finite-size clusters, (ii) solving the problem (exactly if possible) on these clusters (usually by ED), and (iii) treating perturbatively the inter-cluster terms of the Hamiltonian to first order in a strong-coupling expansion.
The core formula resulting from these procedures reads
 \begin{equation}\label{eq:cpt1}
 {\cal G}_{a,b}(Q,\omega) =\left( \frac{{\hat G}(\omega)}{1-{\hat V}(Q){\hat G}(\omega)}\right)_{a,b},
\end{equation}
and
\begin{equation}\label{eq:cpt2}
{\cal G}_{\textrm{CPT}} (q,\omega) = \frac{1}{L} \sum_{a,b=1}^L e^{-iq(a-b)} {\cal G}_{a,b} (Lq,\omega).
 \end{equation}
Here ${\cal G}_{a,b}(Q,\omega)$ is written in a mixed representation of real space indices within the finite-size cluster and Fourier space wavevector between the clusters:
${\hat G}(\omega)$ is the cluster Green's function (computed preferably with open boundary condition~\cite{Potthoff2003}), and ${a,b}$ are the real-space indices for an $L$-site lattice.
The inter-cluster terms are accounted for by ${\hat V}(Q)$ written in the reciprocal superlattice representation.

\begin{figure}[t!]
\includegraphics[width=\columnwidth]{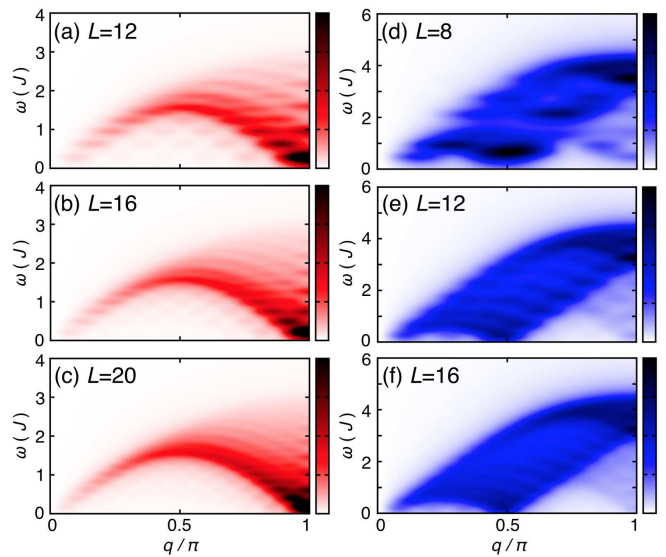}
\caption{
Dynamical structure factors for spin [(a)-(c)] and orbital [(d)-(f)] computed by CPT+ED on lattices of different lengths $L$.
The ED spectra are broadened with a $0.25J$ Lorentzian.
The false color white represents zero intensity, and black represents the maximal intensity [0.4 in (a)-(c) for a pure spin chain; 0.2 in (d)-(f) for the spin-orbital model].
With increasing $L$, the ripple structure resulting from CPT interpolation smooths and the overall spectral shape converges.
}
\label{fig:CPTConvergence}
\end{figure}

\begin{figure}[t!]
\includegraphics[width=\columnwidth]{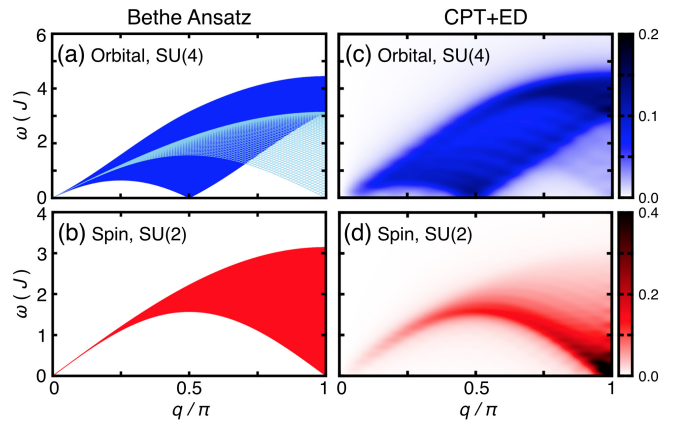}
\caption{
Benchmark of the CPT+ED calculations [(c)-(d)] against exact Bethe-ansatz solutions [(a)-(b)].
The colors in (a) and (b) do not represent the spectral intensity, but only the compact support (nonzero region of a function).
The ED spectra (broadened with a 0.25$J$ Lorentzian) are computed on an $L=16$ and $L=24$ site lattice for (c) and (d), respectively.
The exact spectral shape and overall bandwidth for both the SU(4) spin-orbital model and the SU(2) spin chain are reproduced by CPT+ED at a quantitative level.
}
\label{fig:BetheAnsatz}
\end{figure}

In our case of a one-dimensional (1D) chain, ${\hat V}(Q) = J_{\textrm{eff}}(e^{iQ}\delta_{a,L} \delta_{b,1} + e^{-iQ} \delta_{a,1} \delta_{b,L})$, where $J_{\textrm{eff}}$ is the effective strength of the exchange coupling between inter-cluster spin (or orbital) operators.
We have tried $J_{\textrm{eff}}= J, -J, 0$, and the CPT results only weakly depend on our choice of $J_{\textrm{eff}}$,
as long as $L$ is large enough.
While CPT was originally developed for Hamiltonians {\it without} inter-cluster interactions,
it remains a good approximation of the lattice green function even with the presence of inter-cluster superexchange terms (such as those in the $t-J$ model or the Heisenberg spin chain).
This is because the accuracy of CPT is not directly controlled by including higher order terms in the strong-coupling perturbation theory,
but mainly by increasing the cluster sizes in the simulations~\cite{Senechal2012}.
As shown in Fig. \ref{fig:CPTConvergence}, when $L$ increases,
the overall spectral shape converges quickly for both the spin chain and the spin-orbital model;
the (artificial) ripple structures resulting from CPT interpolation also weaken in intensity and smooth gradually with increasing $L$.

Figure \ref{fig:BetheAnsatz} benchmarks the CPT+ED calculations against the compact supports (regions of nonzero spectral weight) obtained by Bethe-Ansatz solutions.
As seen from the comparison, CPT+ED is capable of reproducing the exact spectral shape and overall bandwidth at a quantitative level.
We note that in these 1D systems, the spin and orbital spectra in the ED calculations already show multiple peaks which spread out widely in energy, implying a fractional nature of the excitations (see Fig. \ref{fig:ChenbosonictJ}).
On the other hand, ED calculations performed on the higher-dimensional counterparts would show only sharp spectral peaks,
and thereby the CPT-interpolated spectra would not display any continuum.

\section*{APPENDIX B: MAPPING THE SPIN-ORBITAL MODEL TO A BOSONIC $t$-$J$ MODEL}
Here we discuss the mapping of the spin-orbital Hamiltonian onto an effective $t-J$ model, which is shown to be valid only when $E_z \ge  E_z^{cr}$.
 
We begin with a more general spin-orbital model:
\begin{align}\label{eq:spinorb_v2}
\mathcal{H}_{\rm gen} &= \sum_{\langle i, j \rangle} \Big( {\bf S}_i \cdot {\bf S}_j + \frac{1}{4} \Big) 
\Big[ J_{ab} \big(T^+_i T^-_j + T^-_i T^+_j \big)  \nonumber \\ 
& + J_b \big( T^z_i - \frac12 \Big) \Big( T^z_j - \frac12 \Big) + 
J_a \Big( T^z_i + \frac12 \Big) \Big( T^z_j + \frac12 \Big) \Big] \nonumber \\
&+ E_z \sum_i T^z_i.
\end{align}
This more realistic spin-orbital model describes systems where the two orbitals $a$ and $b$ under consideration are not equivalent ($e.g.$ $p$ orbitals in alkali hyperoxides~\cite{Wohlfeld2011epl}, or $d-d$ excitations in copper oxides~\cite{Wohlfeld2013}).
When $2 J_{ab} = 2 J_a = 2 J_b \equiv J$, the model is equal to Eq.~(\ref{eq:h}).

Using a similar transformation discussed in Sec. III, we can rewrite Eq. (\ref{eq:spinorb_v2}) as
\begin{align} \label{eq:towardstJ_v2}
\mathcal{H}_{\rm gen} &= J_a \sum_{\langle i, j \rangle} \Big( {\bf S}_{i a} \cdot {\bf S}_{ j a} + \frac{1}{4} \Big) 
+  J_b \sum_{\langle i, j \rangle} \Big( {\bf S}_{i b} \cdot {\bf S}_{ j b} + \frac{1}{4} \Big)  \nonumber \\ 
&+ \frac{J_{ab}}{2} \sum_{\langle i, j \rangle, \sigma, \bar{\sigma}}  \big( {f}_{i a \sigma}^\dagger {f}_{i b \bar{\sigma}} {f}^\dagger_{j b \bar{\sigma}} {f}_{j a \sigma} +
{f}_{i b \sigma}^\dagger {f}_{i a \bar{\sigma}} {f}^\dagger_{j a \bar{\sigma}} {f}_{j b \sigma}  \big) \nonumber \\
&+ \frac12 E_z \sum_{i, \sigma}  \big( {f}_{i a \sigma}^\dagger {f}_{i a \sigma} - {f}_{i b \sigma}^\dagger {f}_{i b \sigma} \big),
\end{align}
where $S_{i a}^+ =  {f}_{i a\uparrow}^\dagger {f}_{i a \downarrow}$ and $S_{i b}^+ = {f}_{i b \uparrow}^\dagger {f}_{i b \downarrow}$.
We then map Eq.~(\ref{eq:towardstJ_v2}) onto a {\it bosonic} $t$-$J$ model [by neglecting interactions between spins in the $a$ orbitals ($ \propto J_a {\bf S}_{i a} \cdot {\bf S}_{ j a}$)] with the following transformations:
\\
(i) We substitute ${f}_{i  a \sigma} = {a}_i f_{i \sigma}$,
where ${a}_i $ fermion carries the orbital degree of freedom, 
and $f_{i \sigma}$ is a spin Schwinger boson subject to the constraint $ {a}^\dagger_i   {a}_i  = \sum_{\sigma}  f^\dagger_{i \sigma}  f_{i \sigma} $.
\\
(ii) We introduce $b_{i \sigma}^\dagger =  {f}^\dagger_{i b \sigma} {a}_{i}$, where $b_{i \sigma}$ are hard-core {\it boson} operators with the constraint $\sum_{\sigma} b^\dagger_{i \sigma} b_{i \sigma} \le 1$.
In this case, the third term in Eq. (\ref{eq:towardstJ_v2}) [denoted as $H_{kin}$] becomes $H_{kin} =  J_{ab} / 2 \sum_{\langle i, j \rangle, \sigma}  [b^\dagger_{i \sigma} b_{j \sigma} (\sum_{\sigma'} f^\dagger_{i \sigma'} f_{j \sigma'}) + h.c.]$.
\\
(iii) We then neglect the $f_{i \sigma}$ Schwinger boson operators (which describe changes of the spin configuration
for electrons in the $a$ orbital), since these do not influence the dynamics of the $b$ bosons.
They do not affect the statistics, either, as the $f$ bosons on different sites commute.

Altogether we arrive at the following Hamiltonian
\begin{align} \label{eq:bosonictJ_v3}
\mathcal{H}_{\rm gen}^{\rm t-J} &=  \frac{J_{ab}}{2} \sum_{\langle i, j \rangle, \sigma}  \big( b^\dagger_{i \sigma} b_{j \sigma} + h.c. \big) 
+J_b \sum_{\langle i, j \rangle} \Big( {\bf S}_{i b} {\bf S}_{ j b} + \frac{1}{4} \Big) \nonumber \\
&+ E_z \sum_{i}  n_{b i},
\end{align}
which is a bosonic $t$--$J$ model~\cite{Boninsegni2001} with $t \equiv  J_{ab}/2$.
In this case, the $a$-orbital electrons can be seen as holes in the spin background formed by electrons in the $b$ orbitals.

We emphasize that the above mapping requires a completely ``silent" spin of electrons in $a$ orbitals,
since in the mapping the interactions $ \propto J_a {\bf S}_{i a} \cdot  {\bf S}_{ j a}$ are neglected (see Fig. \ref{fig:tJcartoon}).
The latter situation occurs when $ J_a=0$, {\it or} when only one electron occupies orbital $a$.
In the spin-orbital Hamiltonian studied in the main text, $J_a$ is finite ($2 J_{ab} = 2 J_a = 2 J_b = J$, as mentioned above).
Therefore, the mapping of the spin-orbital model, Eq.~(\ref{eq:h}), onto an effective $t$-$J$ model,
Eq.~(\ref{eq:bosonictJ_v2}), can be performed only when  $E_z \ge  E_z^{cr}$, 
where a single orbital-flip excitation results in at most one electron in orbital $a$.
In the latter case, substituting $2 J_{ab} = 2 J_b= J$ in Eq.~(\ref{eq:bosonictJ_v3}) renders Eq.~(\ref{eq:bosonictJ_v2}) in the main text.

\bibliographystyle{apsrev4-1}
\bibliography{biblio_orbiton}

\begin{thebibliography}{76}
\bibitem [{ft4()}]{ft4}%
  \BibitemOpen
  \href@noop {} {}\bibinfo {note} {This is the ``reverse{"} transformation with
  respect to the commonly used one for transferring from the Hubbard to the
  Heisenberg model~\cite{Kenzelmann2005}.}\BibitemShut {Stop}%
\end{thebibliography}

\end{document}